# Femtosecond rotational dynamics of $D_2$ molecules in superfluid helium nanodroplets

Junjie Qiang,[1,*] Lianrong Zhou,[1,*] Peifen Lu,[1,†] Kang Lin,[1] Yongzhe Ma,[1] Shengzhe Pan,[1] Chenxu Lu,[1] Wenyu Jiang,[1] Fenghao Sun,[1] Wenbin Zhang,[1] Hui Li,[1] Xiaochun Gong,[1] Ilya Sh. Averbukh,[2] Yehiam Prior,[1,2] Constant A. Schouder,[3] Henrik Stapelfeldt,[3,‡] Igor N. Cherepanov,[4] Mikhail Lemeshko,[4] Wolfgang Jäger,[5] Jian Wu[1,6,7,§]

[1]*State Key Laboratory of Precision Spectroscopy, East China Normal University, Shanghai 200241, China*
[2]*AMOS and Department of Chemical Physics, Weizmann Institute of Science, Rehovot 76100, Israel*
[3]*Department of Chemistry, Aarhus University, Langelandsgade 140, 8000 Aarhus C, Denmark*
[4]*Institute of Science and Technology Austria, Am Campus 1, 3400 Klosterneuburg, Austria*
[5]*Department of Chemistry, University of Alberta, Edmonton, Alberta T6G 2G2, Canada*
[6]*Collaborative Innovation Center of Extreme Optics, Shanxi University, Shanxi 030006, China*
[7]*CAS Center for Excellence in Ultra-intense Laser Science, Shanghai 201800, China*

**Abstract**

Rotational dynamics of $D_2$ molecules inside helium nanodroplets is induced by a moderately intense femtosecond (fs) pump pulse and measured as a function of time by recording the yield of $HeD^+$ ions, created through strong-field dissociative ionization with a delayed fs probe pulse. The yield oscillates with a period of 185 fs, reflecting field-free rotational wave packet dynamics, and the oscillation persists for more than 500 periods. Within the experimental uncertainty, the rotational constant $B_{He}$ of the in-droplet $D_2$ molecule, determined by Fourier analysis, is the same as $B_{gas}$ for an isolated $D_2$ molecule. Our observations show that the $D_2$ molecules inside helium nanodroplets essentially rotate as free $D_2$ molecules.

---

[*] These two authors contributed equally.
[†] pflu@lps.ecnu.edu.cn
[‡] henriks@chem.au.dk
[§] jwu@phy.ecnu.edu.cn

Laser-induced alignment of molecules in the gas phase has been intensively studied in the last decades and used in various applications [1-6]. In particular, moderately intense femtosecond laser pulses can create rotational wave packets, i.e., coherent superpositions of field-free rotational eigenstates, leading to alignment and anti-alignment revivals in narrow, periodically-occurring time windows. For isolated linear and symmetric top molecules, the characteristic revival structure of the time-dependent degree of alignment, explored in a large number of works, persists and does not change provided coupling of the rotational angular momentum to, e.g., the nuclear spin is negligible [7, 8]. On the other hand, if the molecules are in dense gas, collisions with other molecules or atoms will lead to a gradual dephasing of the wave packet and possibly rotational redistribution, which manifests as gradual disappearance of the revival structures [9-11].

Recently, alignment in the short-pulse or nonadiabatic regime (limit) was extended to molecules embedded in liquid helium nanodroplets [12-16]. Studies on OCS, $CS_2$ and $I_2$ molecules showed that when their rotational energy is kept well below the roton energy of the droplet [17, 18], where coupling between rotation and the phonons is weak, revivals are present in the time-dependent degree of alignment traces, reflected as discrete peaks in the corresponding frequency spectra. These observations were interpreted as a consequence of the superfluidity of He droplets. Although the observed rotational dynamics differed strongly from the rotational dynamics of gas-phase molecules, a free-rotor model accounted for the experimental results [15]. It revealed that the difference is due to the smaller effective rotational constant $B_{He}$ and the much larger centrifugal distortion constant $D_{He}$ of the in-droplet molecules [17, 18] compared to those of the gas-phase molecules, $B_{gas}$ and $D_{gas}$, as well as to a distribution rather than a single value of $B_{He}$ and $D_{He}$.

For OCS, $CS_2$ and $I_2$, the $B_{gas}/B_{He}$ ratio lies between 2 and 5, an effect caused by a nonsuperfluid density of He corotating with the molecules [19]. For smaller and lighter molecules, corresponding to larger values of $B_{gas}$, $B_{gas}/B_{He}$ tends towards 1, notably when $B_{gas} > 1$ cm$^{-1}$. This trend, qualitatively explained as the surrounding He atoms not being able to follow the fast-rotating molecules, indicates that the rotation of small molecules inside He droplets should closely resemble that in the gas phase. However, the collected information from a large number of Infrared (IR) spectroscopy studies have established a correlation between $D_{He}$ and $B_{He}$, expressed as $D_{He} = 0.038B_{He}^{1.88}$ (using cm$^{-1}$ units) [18]. If this correlation remains valid for $B_{He} > 1$ cm$^{-1}$, then $D_{He}$ becomes comparable to $B_{He}$, a situation that will lead to rotational dynamics very different from that of gas-phase molecules. On the other hand, the quasiparticle angulon model recently predicted that in the limit of light rotors, $D_{He}$ scales as $B_{He}^{-1}$, thereby deviating from the nearly quadratic dependence [16]. Using the reported analytic formula, we estimate the upper bound of $D_{He} = 0.001$ cm$^{-1}$ for $D_2$ (see Supplemental Material [20], Sec. V). So an obvious question arises: How do small and light molecules ($B_{gas} >> 1$ cm$^{-1}$) rotate inside superfluid He droplets? The purpose of this Letter is to answer this question. It is done by using a non-resonant fs pulse to create rotational wave packets in $D_2$ molecules embedded in helium nanodroplets and measuring the rotational dynamics through timed strong-field ionization. The observed rotational dynamics is essentially the same as that of isolated gas-phase $D_2$ molecules.

A schematic of the experimental setup is given in Fig. 1(a). A continuous helium nanodroplet beam was produced by expanding pre-cooled (16.2 K), high purity (99.9999%), $^4$He gas at 20 bar stagnation pressure through a 5-µm-diameter nozzle. The average size of the droplets is estimated to be 2000 atoms [17]. The helium nanodroplets were doped by passing through a 4-cm-long pick-up cell containing $D_2$ molecules. Two orthogonally-polarized pulsed laser beams were focused by a concave silver mirror ($f$=7.5 cm) onto the $D_2$-doped He droplets in the reaction microscope of a Cold Target Recoil Ion Momentum Spectroscopy (COLTRIMS) setup [21, 22]. The parameters of the pulses in the pump beam, used to induce alignment, were 790 nm, ~40 fs, *y*-polarized, and the parameters of the beam used to probe the alignment, were 395 nm, ~30 fs, *z*-polarized. The peak intensities of the pump and the probe pulses in the interaction region were estimated to be $I_{pump} = 8.0 \times 10^{13}$ W/cm$^2$ and $I_{probe} =$

$2.0 \times 10^{14}$ W/cm$^2$.

In previous studies of laser-induced alignment of molecules in He droplets, the degree of alignment was measured by Coulomb exploding the molecules with an intense laser pulse and recording the emission directions of the fragment ions [12, 23]. Fragment ions like I$^+$, Br$^+$, or S$^+$ from, e.g., I$_2$, C$_6$H$_5$Br and CS$_2$ molecules, respectively, loose some of their initial directionality following the Coulomb explosion event due to scattering on He atoms as the ions move out of the droplet, but their final angular distribution can still be used to determine the degree of alignment [23, 24]. For the much lighter fragment ion D$^+$, the influence of the He scattering on the ion trajectories is more severe, and our experiment shows that the Coulomb explosion probe method is not well-suited (see Supplemental Material [20], Sec. I). Instead, the rotational dynamics is probed by ionizing the D$_2$ molecules with the probe pulse and measuring the dissociative-ionization yield because it depends on the orientation of the molecular axis relative to the polarization of the probe pulse.

The alignment-dependent dissociative ionization process details are illustrated in Fig. 1(b). The probe pulse ionizes and dissociates the D$_2$ molecule via two channels: D$_2 \rightarrow$ D$_2^+ + e \rightarrow$ D$^+$ + D + $e$, denoted as D$_2$(1,0) and D$_2 \rightarrow$ D$_2^+ + e \rightarrow$ D$^+$ + D$^+$ + 2$e$, denoted as D$_2$(1,1). Both channels are sequential, and the first step is multiphoton ionization of D$_2$. The ionization step depends only mildly on the molecular alignment [25, 26]. For the D$_2$(1,0) channel, the second step is a parallel transition between the 1s$\sigma_g$ and 2p$\sigma_u$ states of D$_2^+$ [27], while for the D$_2$(1,1) channel, the second step is dominated by charge-resonance-enhanced ionization of D$_2^+$ [28]. Both of these two processes occur most effectively when the molecular axis aligns along the polarization direction of the laser pulse, i.e., they are alignment-dependent. As a result, the rotational dynamics can be visualized by measuring the time-dependent yield of the dissociative ionization channels [29, 30], i.e., the D$^+$ ion yield, with the highest (lowest) yield expected when the molecules align parallel (perpendicular) to the probe polarization.

The laser pulses interacted not only with D$_2$ molecules inside He droplets but also with free D$_2$ molecules that diffused from the pickup cell to the target region. To eliminate the background contribution from these isolated $D_2$ molecules and obtain a signal that exclusively originates from $D_2$ molecules embedded in the droplets, we recorded $HeD^+$ ions rather than $D^+$ ions. This strategy is similar to that employed in past studies on, e.g., $I_2$ molecules where $HeI^+$ ions were recorded [14, 23]. The $HeD^+$ ions stem from, we believe, $D^+$ fragment ions binding to a He atom as they travel out of the droplet.

Figure 1(c) depicts the measured position y-TOF spectrum of the ions produced by the probe pulse. The spectrum consists of two series of singly charged ionic fragments with mass-to-charge ratio $m/q = 4n$ and $m/q = 4n+2$ ($n = 1,2,3,4…,$ ), which are respectively assigned to $He_n^+$ and $He_nD^+$ ions [31]. The $m/q=4n$ may also have a contribution from $He_{n-1}D_2^+$ ions. The signals with 1000 ns < TOF< 2000 ns are mainly from $D^+$ ions produced when $D_2$ molecules are dissociatively ionized. In what follows, we focus on the peak with $m/q=6$ (marked by the dashed ellipse), which is assigned to $HeD^+$ ions and could only have originated from a $D_2$ molecule embedded in a helium droplet, as discussed above.

Figure 2(a) depicts the yield of the $HeD^+$ ions as a function of the kinetic energy ($E_{kin}$) and the time delay between the pump and the probe pulses. The time interval recorded was from -200 fs to 1000 fs with a step size of 6.7 fs. The $E_{kin}$ axis is on a logarithmic scale, and the yield is normalized for each kinetic energy. Pronounced oscillation structures are visible with an apparent π-phase shift between ions with 0.1 eV < $E_{kin}$ < 1.0 eV and ions with $E_{kin}$ < 0.04 eV. The yields integrated over the two $E_{kin}$ ranges from 100 fs to 1000 fs are plotted in Fig. 2(b). The π-phase shift is evident, and it is seen that the two traces have the same oscillation period of ~185 fs. Here, we focus on the high-$E_{kin}$ fragments, i.e., $HeD^+$, and discuss the low-$E_{kin}$ in Supplemental Material [20], Sec. II.

To explore the further evolution of the yield of the high-$E_{kin}$ ions, measurements were conducted in the interval 1–7 ps with a step size of 10 fs. Figure 3(a) shows that the oscillations in the yield continue essentially without any change, a behavior qualitatively different from the alignment dynamics previously observed for $I_2$, $CS_2$ and OCS embedded in He droplets [15]. The regular sine-like structure of the yield indicates that the $D_2$ rotational dynamics is the result of a wave packet dominated by two rotational quantum states [32]. To investigate if that is the case, the time-dependent $HeD^+$ yield from 1 ps to 7 ps was Fourier transformed. The power spectrum, represented by the red curve in Fig. 3(c), contains three peaks, of which the one centered at 5.35 THz is the dominant one.

Like past studies on both isolated and in-droplet molecules, one may expect that the central value of each spectral peak is given by the frequency of a ($J$–$J+2$) coherence in the rotational wave packets created by the alignment pulse, where $J$ is the rotational angular momentum. Consequently, we assign the three observed peaks at 5.35, 8.92, and 12.40 THz as the (0–2), (1–3), and (2–4) coherences, respectively. The observation that the 5.35 THz peak is by far the strongest shows that this frequency dominates the rotational dynamics. This is consistent with the observed period of ~185 fs in the time-dependent yield of $HeD^+$.

In Fig. 3(d), the central frequencies of the spectral peaks are plotted as a function of $J$. In analogy with previous studies, we fit the data points to $B(4J + 6) - D(8J^3 + 36J^2 + 60J + 36)$, which is the expression for the frequencies of the ($J - J+2$) coherences for a non-rigid linear molecule, characterized by the rotational constant B and the centrifugal distortion constant D [15]. The best fit, shown by the red curve in Fig. 3(d) is obtained for $B_{He} = 29.9 \pm 1.3$ cm$^{-1}$ and $D_{He} = 0.013 \pm 0.060$ cm$^{-1}$, with units converted from THz to cm$^{-1}$ (see the discussion on uncertainty in Supplemental Material [20], Sec. III). There is a large relative uncertainty in both $B_{He}$ and $D_{He}$ because the fit is based only on three points. In particular, for $D_{He}$, the uncertainty is larger than the central value but the result still shows that $D_{He} \ll B_{He}$ and allows us to conclude that the $D_{He} = 0.038 B_{He}^{1.88}$ correlation [18] does not apply to $D_2$ molecules. Moreover, the above mentioned estimate $D_{He} = 0.001$ cm$^{-1}$ lies within the uncertainty range. Our results are the first measured values of $B_{He}$ and $D_{He}$ for $D_2$ molecules embedded in He droplets.

As a reference, the time-dependent yield for isolated $D_2$ molecules in the background, measured using $D^+$ as the observable (blue curve in Fig. 3(b)), exhibits an oscillatory structure similar to that of the $HeD^+$ in Fig. 3(a). The corresponding power spectrum, shown by the blue curve in Fig. 3(c), contains four peaks which we assign as the (0–2), (1–3), (2–4) and (3–5) coherences, respectively. The central positions of the spectral peaks, of which the first three essentially coincide with those for the $D_2$ molecules in He droplets, are plotted as a function of $J$ in Fig. 3(e). Again, the experimental data points are fitted to the non-rigid rotor expression, and we find that the best fit, represented by the blue curve, is obtained for $B_{gas}= 29.98 \pm 0.13$ cm$^{-1}$, $D_{gas}= 0.013 \pm 0.004$ cm$^{-1}$. Within the uncertainty of the experiment, these results are consistent with the values, 29.91 cm$^{-1}$ and 0.01123 cm$^{-1}$ obtained from either experimental spectroscopic measurements or theoretical calculations of the rovibrational spectra of isolated gas-phase $D_2$ molecules [33-35].

Figure 3(c) shows that the relative intensities of the spectral peaks are not the same for the isolated and the in-droplet $D_2$ molecules. As we now explain, this is ascribed to a difference in the initial population, P($J$), of the rotational states. For the isolated $D_2$ molecules, P($J$) is determined by a Boltzmann distribution with a rotational temperature, $T_{rot}$, equal to 295 K and taking into account the 2:1 nuclear spin statistical factor for the even (ortho-$D_2$) and odd (para-$D_2$) states: P(0) = 19%, P(1) = 21%, P(2) = 39%, P(3) = 11%, P(4) = 9%, P(5) = 1%. With this population distribution, the parameters of the experimental alignment pulse, and the $B_{gas}$ and $D_{gas}$ values from the fit of the experimental data, we calculated $\langle \cos^2\theta \rangle(t)$, the standard metric for the degree of alignment, by solving the time-dependent rotational Schrödinger equation; here, $\theta$ is the angle between the alignment pulse polarization and the $D_2$ internuclear axis. Subsequently, Fourier transforming $\langle \cos^2\theta \rangle(t)$ gives a spectrum that is very close to the blue curve in Fig. 3(c) (see Supplemental Material [20], Sec. IV). If P($J$) for the in-droplet $D_2$ molecules is determined by a Boltzmann distribution with $T_{rot}$ = 0.37 K as observed for molecules like $SF_6$ [36], OCS [37], $CS_2$ and $I_2$ [23], then only the $J$ = 0 state should be populated. Only spectral peaks from coherences between states with even $J$ values would be seen. The observation of the (1–3) peak in Fig. 3(c) shows that some of the $D_2$ molecules inside the droplets are initially populated in states with odd $J$. Consequently, the time it takes a He droplet to fly from the pickup cell, where the $D_2$ molecules are embedded, to the interaction region with the laser beam, about 2.2 ms in our experiment, is not sufficient to completely thermalize the rotational state distribution initially at 295 K to one at 0.37 K. We believe this is due to the well-known long time scales for flipping the nuclear spin of the deuterons [38-40].

The information from the time-dependent $HeD^+$ yield and the corresponding power spectrum allows us to conclude that the laser-induced rotational dynamics of the $D_2$ molecules inside He droplets is determined mainly by the (0–2) coherence with frequency 5.35 THz. The unchanged amplitude of $HeD^+$ in the 1–7 ps interval shows that inhomogeneous and homogeneous broadenings are negligible on this time scale. To explore dynamics on longer time scales, data were recorded in the intervals 19.42-20.02 ps and 100.1-100.7 ps, see Figs. 2(c)-(d). Due to the significant acquisition times required, the measurements were restricted to these two selected time windows. Again, the $D^+$ signal from isolated molecules was recorded as a reference. The signals measured in both time intervals demonstrate that the 185-fs oscillations are still present, even after 100 ps, although with a somewhat reduced amplitude compared to the first 7 ps. These observations show that the dominant (0–2) coherence of the rotational wave packet is preserved for more than 500 oscillations and that the $J = 2$ state has a lifetime of at least 100 ps. This is in accordance with a recent theoretical study investigating the influence of the rotational constants, the initial excited rotational state, and the droplet size on the rotational relaxation quantum dynamics of fast rotors inside helium nanodroplets [41]. The calculations predicted that the rotational relaxation of $D_2$ ($J = 2$, $M_J = 0$) in superfluid helium nanodroplets could be particularly slow (> 5 ns) due to the weak molecule-helium interaction. Here, we suggest that the low density of states of superfluid helium [42, 43] at the energies of the excited rotational states of $D_2$ may be responsible for the weak coupling of the molecular rotation with the superfluid helium excitations.

In summary, we experimentally investigated laser-induced field-free molecular alignment dynamics of $D_2$ molecules inside He droplets by measuring the time-dependent yield of $HeD^+$ ions created by ionization with an intense probe laser pulse. The dominant (0–2) rotational coherence has energy (179 cm$^{-1}$) much above the roton energy of the droplets and persists for longer than 100 ps. From the power spectra of the alignment traces, $B_{He}$ is determined and found to be the same as $B_{gas}$. In total, our measurements show that for at least 100 ps, equivalent to > 500 rotational periods, $D_2$ molecules in He droplets rotate as if they were isolated gas-phase particles. This behavior is strikingly different from the in-droplet rotation of all other molecules studied [44]. It would be interesting to investigate the rotational dynamics for significantly longer times to find out when coupling between the $D_2$ rotation and the droplet becomes important. This is in principle possible with our technique. So is studies on other light molecules like HF and $C_2H_2$, which would explore the rotational dynamics of rotors in the gap between the superlight $D_2$ and the heavier species like OCS.

This work is supported by the National Key R&D Program of China (Grant No. 2018YFA0306303); the National Natural Science Fund (Grants Nos. 11834004, 11621404 and 12174109); the project supported by the Shanghai Committee of Science and Technology (Grant No. 19JC1412200). J. Q. acknowledges support by ECNU Cultivation Project of Future Scientist & Outstanding Scholar (Grant No. WLKXJ202003). H. S. acknowledges support from the Villum Foundation through a Villum Investigator Grant No. 25886. M. L. acknowledges support by the European Research Council (ERC) Starting Grant No. 801770 (ANGULON). I. C. acknowledges support by the European Union's Horizon 2020 research and innovation programme under the Marie Sklodowska-Curie Grant Agreement No. 665385.

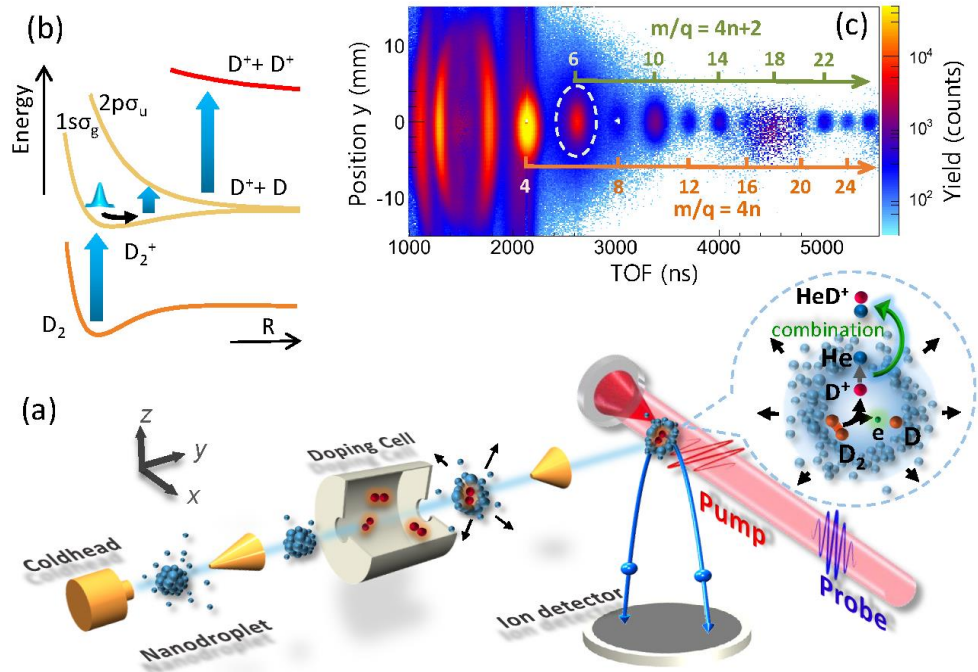

FIG. 1. (a) Schematic diagram of the experimental setup. The pump pulse initiates the rotational dynamics of $D_2$ molecules inside the He nanodroplets, and then the dynamics is measured through dissociative ionization, induced by the probe pulse, and recording of the $HeD^+$ ions. (b) Sketch of relevant potential energy curves of $D_2$, $D_2^+$ and $D_2^{2+}$ (1/R Coulomb potential). The blue vertical arrows illustrate the two involved dissociative ionization channels, i.e., $D_2(1,0)$ and $D_2(1,1)$, see text. (c) Measured position y-TOF spectrum of the ions produced when the probe pulse interacts with the helium nanodroplets doped with $D_2$ molecules. The signal originating from the residual gas in the interaction chamber (e.g., $H_2$, $H_2O$ and $N_2$), recorded when the jet of the pick-up cell was closed, was subtracted from the spectrum for clarity.

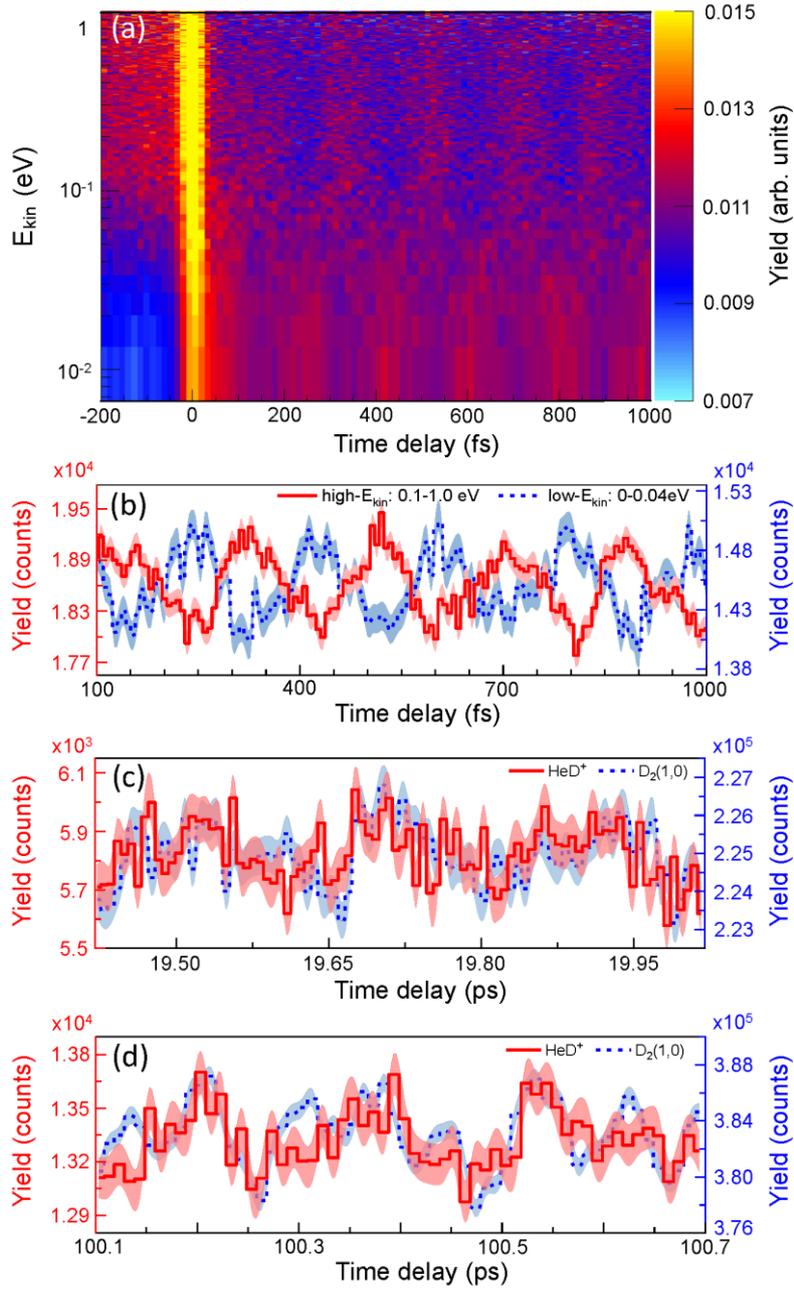

FIG. 2. (a) Yield of the ionic fragments with $m/q=6$ as a function of the kinetic energy and the time delay. (b) Time-dependent yield obtained by integrating the yield over the low-$E_{kin}$ and the high-$E_{kin}$ ranges. (c)-(d) Time-dependent yields of HeD$^+$ (from $D_2$ in He droplets) and D$^+$ (from the $D_2(1,0)$ channel of the isolated gas-phase $D_2$ molecules) in two selected time intervals. The shaded area represents the error bars. The time step sizes for the intervals of 19.42-20.02 ps and 100.1-100.7 ps are 6.7 fs and 10 fs, respectively.

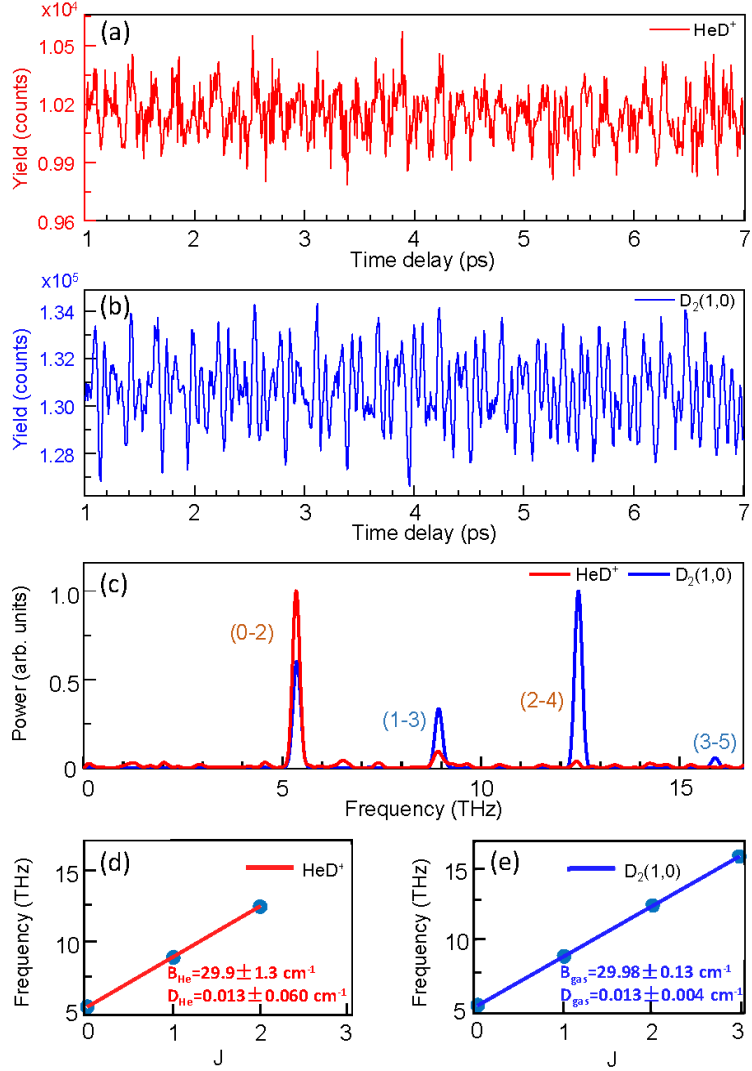

FIG. 3. (a)-(b) Time-dependent yields of HeD$^+$ (from D$_2$ in He droplets) and D$^+$ (from the D$_2$(1,0) channel of the gas-phase D$_2$ molecules) in the 1-7 ps time interval. (c) Power spectra of the yield traces in (a) and (b). (d)-(e) Central frequencies of ($J – J+2$) peaks in the power spectra versus *J*. The full lines represent the best fits using the least square method; see text. The B and D constants from the fits are given with 95% confidence bounds for the in-droplet D$_2$ molecules and the isolated gas-phase D$_2$ molecules.

# Supplemental Material:
# Femtosecond rotational dynamics of $D_2$ molecules in superfluid helium nanodroplets

Junjie Qiang,[1] Lianrong Zhou,[1] Peifen Lu,[1] Kang Lin,[1] Yongzhe Ma,[1] Shengzhe Pan,[1] Chenxu Lu,[1] Wenyu Jiang,[1] Fenghao Sun,[1] Wenbin Zhang,[1] Hui Li,[1] Xiaochun Gong,[1] Ilya Sh. Averbukh,[2] Yehiam Prior,[1,2] Constant A. Schouder,[3] Henrik Stapelfeldt,[3] Igor N. Cherepanov,[4] Mikhail Lemeshko,[4] Wolfgang Jäger,[5] Jian Wu[1,6,7]

[1]*State Key Laboratory of Precision Spectroscopy, East China Normal University, Shanghai 200241, China*
[2]*AMOS and Department of Chemical Physics, Weizmann Institute of Science, Rehovot 76100, Israel*
[3]*Department of Chemistry, Aarhus University, Langelandsgade 140, 8000 Aarhus C, Denmark*
[4]*Institute of Science and Technology Austria, Am Campus 1, 3400 Klosterneuburg, Austria*
[5]*Department of Chemistry, University of Alberta, Edmonton, Alberta T6G 2G2, Canada*
[6]*Collaborative Innovation Center of Extreme Optics, Shanxi University, Shanxi 030006, China*
[7]*CAS Center for Excellence in Ultra-intense Laser Science, Shanghai 201800, China*

## 1. Comparison of the kinetic energy spectra and angular distribution of $D^+$ from isolated $D_2$ and $HeD^+$ from in-droplet $D_2$ molecules.

To illustrate the difference between the fragment ions from ionization of the isolated and the in-helium $D_2$ molecules, we compare the kinetic energy ($E_{kin}$) spectra and angular distributions of the $D^+$ and $HeD^+$ ions. Figure S1(a) depicts the results for the $D^+$ ions, which originate mainly from dissociative ionization of the isolated $D_2$ molecules. The figure shows that at the intensity of the probe pulse, the $D_2(1,0)$ channel, centered at ~1.7 eV, dominates the spectrum. The strong anisotropy in both the angular distributions of both the $D_2(1,0)$ and the $D_2(1,1)$ channels [inset in Fig. S1(a)] is a result of the molecular alignment-dependent dissociative ionization rate. Figure S1(b) depicts the kinetic energy spectrum of the $HeD^+$ ions. We note that the kinetic energies of the $HeD^+$ ions are below 1 eV. This shift of the $E_{kin}$ distribution of the $HeD^+$ to values much lower than those of the $D^+$ ions in the gas-phase spectrum [shown in Fig. S1(a)] is explained by the energy loss of the $D^+$ ions (and of $HeD^+$ ions once they are formed) as they collide with He atoms on their way out of the droplet. The relative loss of $E_{kin}$ is more pronounced than had been observed for heavier iodine fragment ions [1] and is ascribed to a more efficient transfer of kinetic energy in the collisions due to the mass similarity of $D^+$ and He. The anisotropy of the $HeD^+$ angular distribution [see inset in Fig. S1(b)] is less pronounced than that of the $D^+$ ions in the gas phase. Here, too, the observation is ascribed to the fact that collisions between the departing $D^+$ ions and He atoms will significantly change the initial propagation direction of the $D^+$ ions. The reduced anisotropy illustrates why the angular distribution of the $HeD^+$ fragment ions is not a useful observable for characterizing the alignment of $D_2$ molecules in He droplets.

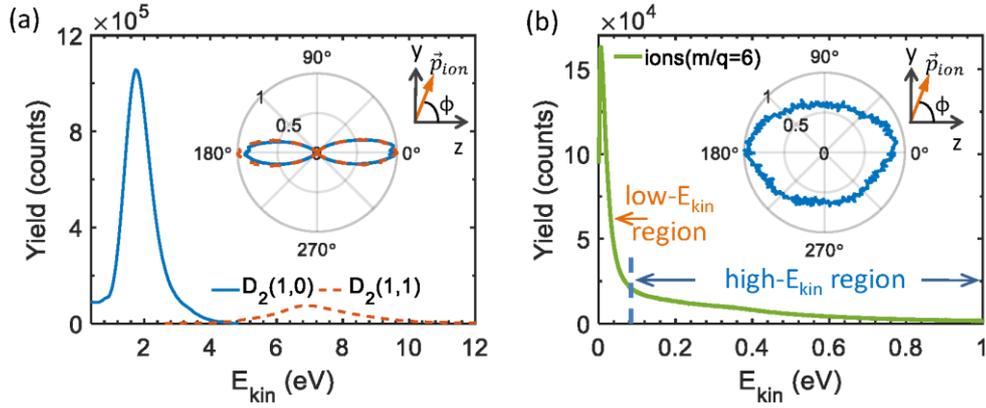

**FIG. S1. (a)-(b) Measured $E_{kin}$ spectra and angular distributions (insets) of the photofragment $D^+$ from the isolated $D_2$ molecules via the $D_2(1,0)$ (solid blue lines) and $D_2(1,1)$ (orange dashed lines) channels and of the ionic fragment $HeD^+$ from the in-droplet $D_2$ molecules. Here, ϕ is the angle of the ejection direction of a fragment ion with respect to the polarization direction of the probe pulse (*z*-axis). The angular distribution in (b) is extracted for the ionic fragment in the high-$E_{kin}$ region.**

## 2. Origin of the low-$E_{kin}$ fragments with *m/q*=6

Here, we discuss the low-$E_{kin}$ fragments with *m/q*=6 assigned to $D_3^+$, as depicted in Fig. S1(b). In our experiments, there was a non-negligible probability that a helium droplet captured two or more $D_2$ molecules when passing through the doping cell. Thus, the fragments with *m/q*=6 may also originate from $D_3^+$, which is produced via an exoergic reaction $D_2^+ + D_2 \rightarrow D_3^+ + D$ [2] or via a dissociative ionization of $D_2$ dimers [3], leaving the droplet as a bare ion. The measured low-$E_{kin}$ fragments in Fig. 2(a) and Fig. S1(b) can be assigned to $D_3^+$ because the momentum of the nascent $D_3^+$ is much smaller compared to the photofragments $D^+$. This is confirmed by the reduced low-$E_{kin}$/high-$E_{kin}$ yield ratio when the pressure of $D_2$ molecules in the pick-up cell is decreased. Although the reactions of $D_3^+$ are supposed to be independent of the molecular alignment, here, the periodical modulation of the time-dependent $D_3^+$ yield is due to the depletion of $D_2^+$ by the alignment-sensitive photofragmentation reactions.

## 3. Determination of the uncertainty of B and D constants from the fit of the experimental data

The frequency resolution of the power spectra ($\Delta f$) is mainly determined by the length of the scan time window (T), i.e., $\Delta f = T^{-1} = 1/6$ ps = 0.167 THz. This frequency resolution, i.e., the smallest frequency division, gives rise to a random error in reading of the central frequencies of the peaks in the power spectra. This random error can usually be one-tenth to one-half of the smallest division. Thus, the random error in reading the central frequencies can range from ±0.02 THz to ±0.08 THz. Accordingly, we read the central frequencies of the peaks in Fig. 3(c) to the same number of decimal places of the random error, i.e., four peaks of 5.37, 8.94, 12.45 and 15.89 THz for the isolated $D_2$ molecule and three peaks of 5.35, 8.92 and 12.40 THz for the in-droplet $D_2$ molecule. Finally, the B and D constants with 95% confidence bounds are obtained by fitting these data points [blue circle, in Figs. 3(d)-(e)], to $B(4J + 6) - D(8J^3 + 36J^2 + 60J + 36)$ using the least square method. For the isolated $D_2$ molecule, the best fit is obtained for $B_{gas}$ = 29.98 ± 0.13 cm$^{-1}$ and $D_{gas}$ = 0.013 ± 0.004 cm$^{-1}$, and for the in-droplet $D_2$ molecule, the best fit is obtained for $B_{He}$ = 29.9 ± 1.3 cm$^{-1}$ and $D_{He}$ = 0.013 ± 0.060 cm$^{-1}$.

## 4. Simulated and experimental results for isolated $D_2$ molecules

We calculated the degree of alignment $\langle \cos^2 \theta \rangle (t)$ by solving the time-dependent rotational Schrödinger equation using the $B_{gas}$ and $D_{gas}$ values from the fit of the experimental data. Here, $\theta$ is the angle between the alignment pulse polarization direction and the $D_2$ internuclear axis. The calculation was averaged over the initially populated rotational states, given by a Boltzmann distribution at T=295 K. In our experiments, the probe pulse (*z*-polarized) is perpendicularly polarized to the pump pulse (*y*-polarized). As a result, the time-dependent $D^+$ yield for isolated $D_2$ molecules is expected to be proportional to the degree of alignment along the *z*-direction. For a better comparison between the calculated and the experimental results, we plot $\frac{1}{2}\left(1 - \langle \cos^2 \theta \rangle (t)\right)$ (dotted blue curve) in Fig. S2(a), which has a strong resemblance with the measured time-dependent yield of $D^+$ from the $D_2(1,0)$ channel of isolated gas-phase $D_2$ molecules (solid red curve). Figure S2(b) depicts the corresponding power spectra of the time-dependent traces in Fig. S2(a). The good agreement between the experimental and calculated results validates the probe method used in our experiment.

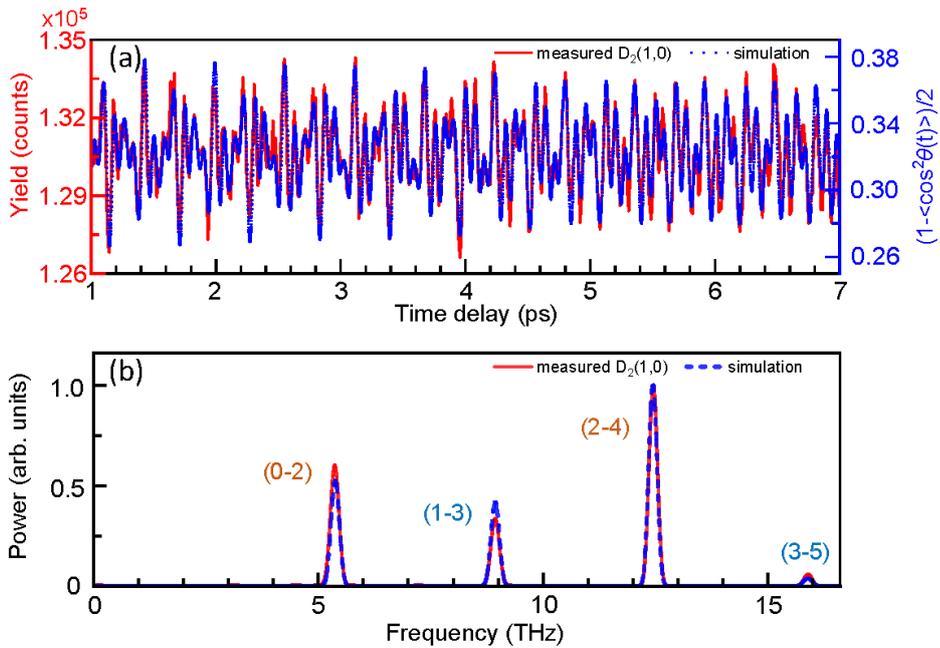

**FIG. S2. (a) Calculated degree of alignment along the polarization direction of the probe pulse for $D_2$ molecules at T=295 K and measured time-dependent yield**

trace of $D^+$ from $D_2(1,0)$ of isolated $D_2$ molecules. (b) Power spectra corresponding to the two traces in (a).

## 5. Derivation of the estimate for $D_{He}$

We use the refined expressions for effective spectroscopic constants obtained within the simplified angulon model in the limit of light rotors [4]

$$\frac{B_{He}}{B} = 1 - \frac{2u^2}{\Delta_0^2}; \quad D_{He} = \frac{4B^2u^2}{\Delta_0^3}$$

, where u characterizes the anisotropy of the molecule-helium interaction, $\Delta_0 = \omega + B\lambda(\lambda + 1)$ with $\omega = 6$ cm$^{-1}$ and $\lambda = 2$ being the roton energy and the dominating anisotropic component of the $D_2$-He potential energy surface (PES), respectively. Inspecting the $H_2$-He PES [5], we estimated $u = 1.3$ cm$^{-1}$ as defined in Ref. [6]. Substituting the parameters into the equation, we obtain $D_{He} = 0.001$ cm$^{-1}$ and negligible change in B of 0.003 cm$^{-1}$.